\documentclass{article}

\usepackage{microtype}
\usepackage{graphicx}
\usepackage{subfigure}
\usepackage{booktabs} 

\usepackage{hyperref}

\usepackage[accepted]{icml2020}

\icmltitlerunning{Ensuring CIA of Deployed ML Systems}

\begin{document}

\twocolumn[
\icmltitle{Green Lighting ML: Confidentiality, Integrity, and Availability\\ 
of Machine Learning Systems in Deployment}



\icmlsetsymbol{equal}{*}

\begin{icmlauthorlist}
\icmlauthor{Abhishek Gupta}{equal,maiei,ms}
\icmlauthor{Erick Galinkin}{equal,maiei,r7}
\icmlcorrespondingauthor{Montreal AI Ethics Institute}{abishek@montrealethics.ai and erick@montrealethics.ai}
\end{icmlauthorlist}

\icmlaffiliation{maiei}{Montreal AI Ethics Institute, Montreal, Quebec, Canada}
\icmlaffiliation{ms}{Microsoft Corporation, Redmond, WA, USA}
\icmlaffiliation{r7}{Rapid7, Boston, MA, USA}

\icmlkeywords{Machine Learning security, MLOps, ML in production, adversarial attacks, trustworthy AI}

\vskip 0.3in
]



\printAffiliationsAndNotice{\icmlEqualContribution} 

\begin{abstract}
Security and ethics are both core to ensuring that a machine learning system can be trusted.
In production machine learning, there is generally a hand-off from those who build a model to those who deploy a model.
In this hand-off, the engineers responsible for model deployment are often not privy to the details of the model and thus, the potential vulnerabilities associated with its usage, exposure, or compromise.
Techniques such as model theft, model inversion, or model misuse may not be considered in model deployment, and so it is incumbent upon data scientists and machine learning engineers to understand these potential risks so they can communicate them to the engineers deploying and hosting their models.
This is an open problem in the machine learning community and in order to help alleviate this issue, automated systems for validating privacy and security of models need to be developed, which will help to lower the burden of implementing these hand-offs and increasing the ubiquity of their adoption.
\end{abstract}

\section{Current Landscape}
Today, there are well-understood frameworks both for detailing model characteristics~\cite{mitchell2019model, arnold2019fact} and documenting datasets~\cite{gebru2018datasheets, holland2018dataset}.
These are widely cited as mechanisms to bring more transparency, auditing, and trust to machine learning.
However, adoption has lagged behind and the open problem is why adoption has failed to catch on.

We hypothesize that the reason is that these pieces of documentation are too onerous to create in practice. 
Specifically, creation of this sort of documentation is a highly manual process and even in cases where these documents are produced, the value is quite small given the lack of standardization, practical value, and fragmented understanding of the utility of such documentation~\cite{thomas2001documentation}.
People are not likely to create documentation for documentation's sake, especially if there is no enforcement mechanism, so we must consider automated processes that help to ease the burden and are deeply integrated into existing development and deployment workflows.

As an illustrative example - many machine learning systems are vulnerable to adversarial examples.
Although awareness is relatively high due to the high profile of some research published on the topic~\cite{athalye2017synthesizing, carlini2017towards, athalye2018obfuscated} there remains limited adoption of techniques to mitigate these issues.
Moreover, very few model APIs incorporate meaningful input validation or sanitization other than verifying that the input has valid dimensions.
This can lead to exploitation of vulnerabilities in machine learning frameworks~\cite{stevens2017summoning}.
The risks to models need to be better understood by the data scientists and machine learning engineers who build the models, as well as the engineers who deploy the models into production.

\section{Continuous Integration of Ethical Principles}
\label{ci}
We propose that integration of ethical and security principles via MLFlow~\footnote{http://mlflow.org} or some other automated tool into the ML development lifecycle can ease adoption and ensure these same principles are put into practice rather than merely discussed in hypothetical scenarios.
Existing tools like Deon~\footnote{https://deon.drivendata.org/} allow for easy and semi-automated checking of ethical concerns via configurable mechanisms including data storage, modeling, and deployment.
Often in DevOps, it is these ``blue lights and green boxes'' from CI/CD tools that are looked for to ensure that a product can be trusted for use \cite{duvall2007continuous} rather than an analysis of the associated documentation which has a higher burden in terms of resources and time required to parse them.
We predict that similar principles would hold in ML deployments whereby indicators with low cognitive load requirements will help to guide developers and consumers on the trustworthiness of those systems when it comes to vulnerability to adversarial attacks. 
Just as higher trust is placed in those open-source repositories that have badges with passing build status, high code-coverage, and more \cite{trockman2018adding}, we posit that such indicators will usher a focus on ML systems that emphasize these practices.

\section{Vulnerabilities in Model Development}
Vulnerabilities in model development occur when a creator does not work to mitigate bugs in the underlying structure of the model. 
This can be quite challenging as so often, the vulnerabilities are: 
\begin{itemize}
    \itemsep 0em
	\item Unknown
	\item Difficult to test for
	\item Difficult to exploit
\end{itemize}
If we automate the process of testing for these bugs or can use an algorithmic impact assessment~\cite{calvo2020advancing} to determine the potential risks associated with deployment and use, then better decisions can be made about whether some defense needs to be built into the model or the surrounding software infrastructure. 
This extends beyond traditional cybersecurity practices which are often woefully underprepared for the new attack surfaces that are opened up due to the integration of machine learning into the system.
To wit, researchers at Microsoft~\cite{Siva_Kumar_2020} found that in a survey of 28 companies, including 10 cybersecurity companies, 22 of them did nothing to secure their ML systems, demonstrating this endemic problem.

\section{Vulnerabilities in Model Deployment}
Many of the model vulnerabilities today are not actually incurred in development - they occur at deployment time.
Much of data science is performed in notebooks, and there is a real reproducibility crisis, especially in the way how notebooks are used~\cite{lyu1996handbook}.
When it comes time to deploy a model, those in charge of infrastructure remain focused on uptime, availability, and scalability \cite{lyu2007software}.
Though infrastructure engineers are aware of general security best practices, the unique threat landscape of machine learning is often alien to them, and even to security practitioners~\cite{kumar2020adversarial}.

\section{Mind the Gap}
Since adoption of existing frameworks is limited due to implementation friction, we have a gap between the ideals of deploying ethical, robust, and trustworthy ML and the practical reality of deploying ML systems.
There are very few concrete standards to which ML systems adhere, and integrating those into the development processes can help move us toward these goals.
To move from theory into practice, we need to build a framework that allows for seamless integration into the design, development, and deployment workflows which will:
\begin{itemize}
    \itemsep 0em
	\item achieve ubiquity in adoption
	\item create standards which allow for cross-implementation comparisons
	\item evoke community-driven collaboration to build up security best practices in this domain
\end{itemize}

\section{Future Research}
Moving forward, we will need to build a prototype of this applied framework and test it with beachhead organizations to gather evidence on the efficacy of the approach.
We believe that there is potential to leverage existing frameworks such as the ones mentioned in \autoref{ci} especially such that it reduces the friction of integration and adoption of completely new tools for this purpose. 

A method of ``risk scoring'' for models will need to be developed, which will also require standardization of definitions across the machine learning security community.
Some preliminary work has been done on creating an exposure metric for unintended memorization in neural networks~\cite{carlini2018secret}, but the focus there is extremely limited and a broader concept of risk scoring is needed to ensure that deployed ML systems are protected from adversarial attacks, model theft~\cite{tramr2016stealing}, and model inversion~\cite{fredrikson2015model}.

\section{Conclusion}
Development and deployment of secure, trustworthy models is an open problem which plagues the machine learning community.
Current methods for ensuring the security and trust of models are too onerous to ensure adoption, leading to the current gap between ideation and adoption.
Defining roles and responsibilities for safeguarding the confidentiality of data along with the integrity and availability of models will be crucial for solving this problem and creating robust, ethical, and trustworthy production-grade models.
In order to facilitate this, a seamless framework, integrated into existing development and deployment workflows, for conducting risk assessments must be developed to ease adoption.

%
%

\bibliography{example_paper}
\bibliographystyle{icml2020}

\end{document}